\documentclass[preprintnumbers,pre,twocolumn]{revtex4}

\usepackage{amsmath}
\usepackage{graphicx,caption,subcaption,float}
\usepackage[symbol]{footmisc}
\begin{document}

\author{Korosh Mahmoodi$^{1}$\footnote{Corresponding author: free3@yahoo.com}}
\author{Bruce J. West$^{2}$}
\author{Paolo Grigolini$^{1}$}
\affiliation{1) Center for Nonlinear Science, University of North Texas, P.O. Box 311427,
Denton, Texas 76203-1427, USA\\
2) Information Sciences Directorate, Army Research Office, Research Triangle
Park, NC }
{First submission: PRE 26Oct2016  }

\begin{abstract}
The crucial aspect of this demonstration is
the discovery of renewal events, hidden in the computed dynamics of a
multifractal metronome, which enables the replacement of the phenomenon
of strong anticipation with a time delayed cross-correlation between the driven and the
driving metronome. We establish that the phenomenon of complexity matching, which is the theme
of an increasing number of research groups, has two distinct measures. One
measure is the sensitivity of a complex system to environmental
multi-fractality; another is the level of information transfer, between two
complex networks at criticality.  The cross-correlation function is evaluated in the ergodic long-time limit, but its delayed maximal value is  the signature of information 
transfer occurring in the non ergodic short-time  regime. It is shown that a more complex system transfers its multifractality to a less complex system while the reverse case is not possible. 

\

\textit{Keywords\/{: Transfer of Multifractality, Complexity matching, Crucial events, Ergodicty breaking, Multifractal metronome, Multifractal Decision Making model }}

\end{abstract}

\title{On the Dynamical Foundation of Multifractality}
\maketitle

\section{Introduction}

The central role of complexity in understanding nonlinear dynamic phenomena
has become increasingly evident over the last quarter century, whether the
scientific focus is on ice melting across the car windshield, stock prices
plummeting in a market crash, or the swarming of insects \cite{west}. It is
remarkable, given its importance in modulating the behavior of dynamic
phenomena, from the cooperative behavior observed in herding, schooling, and
consensus building, to the behavior of the lone individual responding to
motor control response tasks, that defining complexity has been so elusive.
As Deligni\`{e}res and Marmelat \cite{delignieres12} point out, a complex
system consists of a large number of infinitely entangled elements, which
cannot be decomposed into elementary components. They go on to provide an
elegant, if truncated, historical review of complexity, along with its
modern connection to fluctuations and randomness. They provide a working
definition for complexity, as done earlier by West \textit{et al }\cite{west}, as a balance between regularity and randomness.

\subsection{Complexity management} \label{spectrum}

In order to sidestep the impasse of providing an absolute definition of
complexity, West \textit{et al}. \cite{west} introduced the \textit{complexity matching effect} (CME). This effect details how one complex
network responds to an excitation by a second complex network, as a function
of the mismatch in the measures of complexity of the two networks. An
erratic time series generated by a complex network hosts crucial events,  namely events characterized by the following 
properties.  The time distance $\tau$ between two consecutive events has a probability density
function (PDF), $\psi(\tau)$,  that is inverse power law (IPL) with IPL index $\mu < 3$. Different pairs of consecutive times correspond to time distances with no correlation, renewal property. The measure of complexity
is taken to be the IPL index $\mu$. Aquino \textit{et al.} \cite{aquino10} show that a
complex network S with IPL-index $\mu <2$, has no characteristic time
scale, and in the long-time limit is insensitive to a perturbation by a
complex network with a finite time scale, including one having oscillatory
dynamics. It is important to notice that when $\mu < 2$, the first moment of $\psi(\tau)$ is divergent, thereby generating a condition of perennial aging, while the condition $\mu < 3$, making the second moment of $\psi(\tau)$ divergent, generates non-stationary fluctuations that become stationary in the long-time limit. The spectrum of these fluctuations
in the region $2 < \mu < 3$ can be evaluated using the ordinary Wiener-Khintchine theorem, whereas the region of perennial aging, $\mu < 2$, requires the 
adoption of a generalized version of this theorem \cite{mirko}.

The network S is expected to be sensitive to perturbations having
the same IPL index. This observation generated the plausible conjecture that
a complex network, with a given temporal complexity, is especially sensitive
to the influence of a network with the same temporal complexity, this being,
in fact, a manifestation of CME.

The crucial events are generated by complex systems at criticality \cite{original1}. 
On the basis of this property, Turalska \textit{et al}. \cite{synchronization} afforded strong numerical
support to the CME, showing that a network at criticality
is maximally sensitive to the influence of a identical network also at
criticality. In this case, the IPL-index of both the perturbed network, $\mu
_{S}$, and the IPL-index of the perturbing network, $\mu _{P}$, must be
identical, because the perturbing and the perturbed network are identical
systems in the same condition, that being the criticality condition. It is
known \cite{beig} that at criticality $\mu _{S}=\mu _{P}=1.5$, thereby
implying that the two systems share the same temporal complexity, with the
same lack of a finite time scale. The CME was subsequently generalized to
the \textit{principle of complexity management} (PCM), where the network
response was determined when both the perturbing and responding networks
have indices in the interval $1<\mu <3$. This latter condition was studied
by means of ensemble averages \cite{aquino10,aquino11} and time averages 
\cite{piccinini16}, leading to the discovery that in the region of perennial
aging $1<\mu <2$ the evaluation of complexity management requires special
treatment. This conclusion is based on the observation that renewal events
are responsible for perennial aging, with the occurrence of the renewal
events in the perturbed network being affected by the occurrence of the
renewal events in the perturbing network. A careless treatment, ignoring this condition may lead to misleading observations characterized erratic behavior making it impossible to realize the correlation between the perturbed and the perturbing 
signal \cite{piccinini16}. This is where the adoption of 
the multi-fractal perspective adopted by the authors of \cite{delignieres} to study CME may turn out to be
more convenient that the adoption of the method of cross-correlation functions.

There is a growing literature devoted to the interpretation that $1/f$-noise
is the signature of complexity, where the spectra of complex phenomena are
given by $1/f^{\nu }$. The complexity in time series is generically called $1/f$-noise or $1/f$-fluctuations, even though empirically the IPL index lies
in the interval $0.5<\nu <1.5$. The $1/f$-behavior can be detected by
converting the underlying time series into a diffusion process. This
conversion of data allows us to determine the corresponding Hurst exponent 
$H $ for the diffusion process, which is well known to be related to the
dimension of fractal fluctuations \cite{mandelbrot77}. Consequently, we
obtain for the scaling index of the spectrum
\begin{equation} \label{mandelbrot}
\nu =2H-1,
\end{equation}
It is important to remark that this approach rests on the Gaussian assumption requiring some caution when dynamical complexity is incompatible with
the Gaussian condition. It has been observed \cite{scafetta} that the condition $\mu > 2$  generates a diffusion process that, interpreted as Gaussian,
yield the Hurst scaling
\begin{equation} \label{scaling}
H = \frac{4-\mu}{2},
\end{equation}
which plugged into Eq. (\ref{mandelbrot})
yields
\begin{equation}
\nu =3-\mu .  \label{nu}
\end{equation}
Eq. (\ref{nu}) is valid also for $\mu < 2$, but in this case it requires a theoretical derivation taking into explicit account the condition of perennial aging \cite{mirko}. 
It is important to stress that $\nu > 1$ is consequently a sign of the action of crucial events. 
The spectrum of a complex network at criticality with crucial events with IPL index $\mu < 2$, expressed in terns of the frequency $\omega = 2\pi f$, is \cite{mirko}:
\begin{equation} \label{mirkospectrum}
S(\omega) \propto \frac{1}{L^{2-\mu}} \frac{1}{\omega^{3-\mu}},
\end{equation}
where $L$ is the length of the observed time series.  In this paper we shall use this expression  to prove that the multi-fractal  metronome used by the authors of Ref. \cite{delignieres}
is driven by crucial events responsible for CME \cite{west}  and complexity management PCM \cite{aquino10}.

Fractal statistics appear to be ubiquitous in time series characterizing
complex phenomena. Some empirical evidence for the existence of $1/f$-noise within the brain
and how it relates to the transfer of information, helps set the stage for
the theoretical arguments given below. The brain has been shown to be more
sensitive to $1/f$-noise than to white noise \cite{soma03}; neurons in the
primary visual cortex exhibit higher coding efficiency and information
transmission rates for $1/f$-signals than for white noise \cite{yu05}; human
EEG activity is characterized by changing patterns and these fluctuations
generate renewal events \cite{gong07}; reaction time to stimuli reveals that
the more challenging the task, the weaker the cognitive $1/f$-noise produced 
\cite{correll08}. Of course, we could extend this list of brain-related
experiments, or shift our attention to other complex systems, but the point
has been made.  

Here we stress that according to the analysis of \cite{gemignani} the brain dynamics 
is a source of ideal $1/f$-noise being characterized by crucial events with
$\mu \approx 2$ in accordance with an independent observation made by Buiatti \emph{et al} \cite{buiatti} of IPL index $\mu$ ranging from
$1.7$ to $2.3$. It is interesting to notice that heartbeat dynamics of healthy patients were proved to host crucial events with $\mu$ close to the ideal condition
$\mu = 2 $ \cite{heartbeat} and that the recent work of Bohara \emph{et al} \cite{gyanendra}, in addition  to confirming this observation  lends support to interpreting the meditation-activated brain-heartbeat synchronicity \cite{meditation} as 
a form of CME  of the same kind as that observed by Deligni\'{e}res and co-workers \cite{delignieres}, namely, as a transfer of multi-fractality.

\subsection{Experiments, multi-fractality and ergodicty breaking}

With the increase in geopolitical tensions between states, the enhanced
social media connectivity between individuals along with the rapid progress
of neurophysiology, societal behavior has become one of today's more
important scientific topics, forcing researchers to develop and adopt new
interdisciplinary approaches to understanding. As pointed out by Pentland 
\cite{pentland}, even the most fundamental social interaction, that being
the dialogue between two individuals, is a societal behavior involving
psychology, sociology, information science and neurophysiology.
Consequently, we try and keep the theoretical discussion outside any one
particular discipline and focus our remarks on what may apply across
disciplines.

Consequently, the same issue of dyadic interaction can be studied from the
theoretical perspective of the Science of Complexity, which we interpret as
an attempt to establish a fruitful interdisciplinary perspective. This view
recognizes that the interaction between phenomena from different disciplines
requires the transfer of information from one complex system to another. The
adoption of this interdisciplinary perspective naturally leads us to use the
previously introduced notion of \emph{complexity matching} \cite{kello,delignieres12,west}. On the other hand, the transition in modeling
from physics to biology, ecology, or sociology, gives rise to doubts
concerning the adoption of the usual reductionism strategy and establishes a
periodicity constraint that is often ignored by physical theories \cite{iberall}.

The observation of the dynamics of single molecules \cite{singlemolecules},
yields the surprising result that in biological systems the ergodic
assumption is violated \cite{chemphyschem}. On the basis of real
psychological experiments, for instance the remarkable report on the
response of the brain to the influence of a multi-fractal metronome \cite%
{stephen}, complexity matching has been interpreted as the transfer of a
scaling PDF from a stimulus to the brain of a stimulated subject. More
recent experimental results \cite{delignieres} confirm this interpretation,
based on the transfer of global properties from one complex network to
another, termed \emph{genuine complexity matching}, while affording
suggestions on how to distinguish it from more conventional local discrete
coupling.

We notice that the PCM \cite{west} relies on the crucial role of
criticality and ergodicity breaking, in full agreement with the concept of
the transport of global properties from one complex network to another. This
agreement suggests, however, that a connection exists, but has not yet been
established, between multi-fractality and ergodicity breaking, in spite of
the fact that current approaches to detecting multi-fractality are based on
the ergodicity assumption \cite{challenge}. Note that multi-fractality is
defined by a time series having a spectrum of Hurst exponents (fractal
dimensions), which is to say the scaling index changes over time \cite%
{feder88}, resulting in no single fractal dimension,\ or scaling parameter,
characterizing the process. Instead there is a unimodel distribution of
scale indices centered on the average Hurst exponent. In psycho-physical
experiments, for example, a person is asked to synchronize a tapping finger
in response to a chaotic auditory stimulus, and complexity matching is
interpreted as the transfer of scaling of the fractal statistical behavior
of the stimulus, to the fractal statistical response of the stimulated
subject's brain. This response of the brain, in such a motor control task,
when the stimulus is a multifractal metronome, has been established \cite{stephen}.

\begin{figure}
\includegraphics[width=0.5\textwidth]{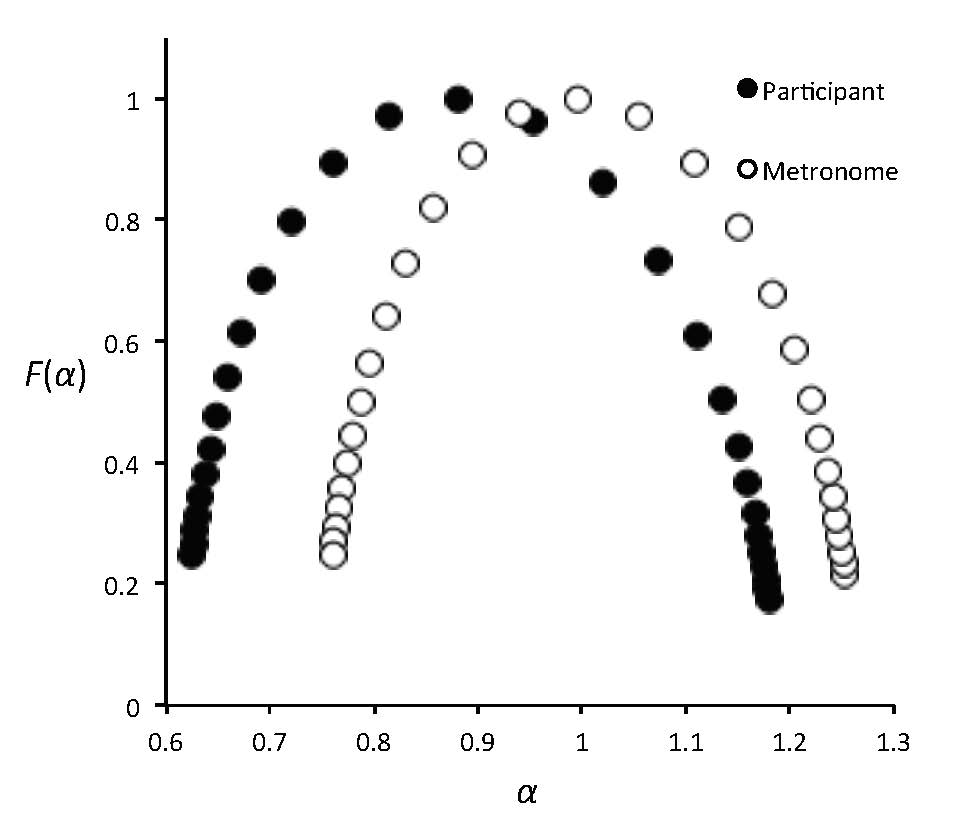}
\caption{Multifractal spectra  for  the  participant  (black  circles)  and  the  metronome  (white  circle). This figure is derived with permission from the left panel of Fig. 3 of Ref. \cite{delignieres}}
\label{bruce.jpg}
\end{figure}

Experimental results \cite{delignieres} confirm this interpretation, based
on the transfer of global properties from one complex network to another. In
these latter experiments the multifractal metronome generates a spectrum of
fractal dimensions $f(\alpha )$ as a function of the average singularity
strength of the excititory signal and it is this dimensional spectrum that
is captured by the brain response simulation, as depicted in Figure \ref{bruce.jpg}. The multi-fractal behavior manifest by the unimodal
distribution provides a unique measure of complexity of the underlying
network.It is worth noting that the same displacement of the metronome
spectrum, from the body response spectrum, is observed for walking in
response to a multifractal metronome.

The main purpose of the present article is to establish that the
multi-fractal arguments advanced by many advocates of complexity matching
must be compatible with data displaying ergodicity breaking. The dynamical
model that we use to establish this connection is the multi-fractal
metronome \cite{stephen}, described by the periodically driven rate equation
with delay: 

\begin{equation}
\dot{x}=-\gamma x(t)-\beta sin\left( x(t-\tau _{m})\right) .
\label{metronome}
\end{equation}

This model, originally introduced by Ikeda and co-workers \cite%
{ikeda1,ikeda2}, was adopted by Voss \cite{voss} to illustrate the
phenomenon of anticipating synchronization. In the present article, we adopt
Eq. (\ref{metronome}) to mimic the output of a complex network, generating
both temporal complexity \cite{temporalcomplexity} and periodicity \cite{zare}. More precisely, setting $\gamma =1$, we use Eq. (\ref{metronome})
with only two adjustable parameters; the amplitude of the sinusoidal driver $\beta$ and the delay time $\tau _{m}$, to determine the nonlinear dynamics
of a metronome with a multi-fractal time series. We should use this simple
equation to study the joint action of renewal events and periodicity, which
may have the effect of either annihilating or reducing the renewal nature of
events. This is, however, a challenging issue that we plan to discuss in
future work. In this paper, on purpose, we establish a time separation
between temporal complexity and periodicity and we establish the accuracy of
this separation by means of the aging experiment.

We show that this simple equation, with a careful choice of the parameters 
$\beta $ and $\tau _{m}$, generates the same temporal complexity as that
produced at criticality by a large number of interacting units in a complex
network \cite{beig}, yielding ergodicity breaking. This earlier work shows
that in the long-time regime the nonlinear Langevin equation, yielding
ergodicity breaking in the short-time regime, becomes equivalent to an
ordinary linear Langevin equation. Herein we use this theory to establish the cross-correlation between a perturbed network $S$ and a perturbing network $P$, identical to the network
$S$, when both networks are at criticality.  Using the important property that in the long-time limit 
the non-linear Langevin equation becomes identical to an ordinary Langevin equation,
we find an exact analytical expression for the cross-correlation between $S$ and $P$. We prove that the mullti-fractal metronome, with a suitable choice of the parameters 
$\gamma$, $\beta$ and $\tau_m$ generates crucial events and that the temporal complexity of these events is identical to that a complex network at criticality, more precisely 
the complex network of Section \ref{dmm}. On the basis of this equivalence we predict that  the cross-correlation function between the metronome equivalent to the network $S$ and the 
metronome equivalent to the metronome $P$ should be identical to the analytical cross-correlation function  between network $S$  and network $P$.  
This prediction is supported with a surprising accuracy by the numerical results of this paper.

The strategy of replacing the output of a complex network with the time
series solution to a multi-fractal metronome equation yields the additional
benefit of avoiding numerically integrating the equations of motion for a
complex dynamic network, involving the interactions among a large number of
particles in a physical model, people in a social model, or neurons in a
model of the brain. Although the interplay between complexity and
periodicity is an issue of fundamental importance \cite{zare}, herein we
focus on the IPL complexity, hidden within the dynamics of Eq. (\ref{metronome}); complexity that was overlooked in earlier research on this
subject.

As pointed out earlier, with our arguments about the generalization of the Wiener-Khintchine theorem in the case of perennial aging we prove numerically 
that the spectrum $S(\omega)$ of the metronome fits very well the prediction of Eq. (\ref{mirkospectrum}). On the other hand, we prove directly that a network at criticality is the source of the broad multi-fractal spectrum used in Ref. \cite{delignieres} to discuss CME.

\subsection{Outline of paper}

The outline of the paper is as follows. Section II shows that a complex
network at criticality generates a distinct multi-fractal spectrum. We
devote Section III to detecting the renewal events hidden within the
dynamics of Eq.(\ref{metronome}) and we establish the equivalence between the
multi-fractal metronome and a complex network at criticality. In Section IV
we find the analytical expression for the cross-correlation between two
complex networks at criticality and we prove numerically that the two
equivalent multi-fractal metronomes generate the same cross-correlation.
Section V illustrates the transfer of the multi-fractal spectrum from a
complex to a deterministic metronome and Section VI is devoted to concluding
remarks.

\section{Criticality,  Decision Making Model and multifractality} \label{dmm}

\subsection {Multifractality of the Decision Making Model}

As an example of criticality we adopt the Decision Making Model (DMM) widely illustrated in the earlier work of Refs. \cite{original, synchronization, beig, second,third,fourth,fifth}. For a review please consult the book of Ref. \cite{20}. To clarify the connection between criticality and multi-fractality, we 
use the DMM. This model rests on a network of N units that have to make a choice between two states, called $C$ and $D$. The state $C$ corresponds to the value $\xi = 1$ and the state $D$ corresponds to the value $\xi= -1$.  The transition rate from $C$   to $D$,  $g_{CD}$, is given by
\begin{equation}
g_{CD} =  g_0 exp \left[ - K \left(\frac{M_C - M_D}{M}\right) \right]
\end{equation}
and  the transition rate from $D$  to the $C$, $g_{DC}$, is given by
\begin{equation}
g_{DC} = g_0 exp \left[  K \left(\frac{M_C - M_D}{M}\right)  \right].
\end{equation}
The meaning of this prescription is as follows. The parameter $1/g_0$ defines a dynamic time scale and we set $g_0 = 0.1$ throughout.  Each individual has $M$ neighbors (four in the case of the regular two-dimensional lattice used in this article). The cooperation state is indicated by $C$ and the defection state 
by $D$. If an individual is in  $C$, and the majority of its neighbors 
are in the same state, then the transition rate becomes smaller and the individual sojourns in the cooperation state for a longer time. If the majority of its neighbors
are in $D$, then the individual sojourns in the cooperator state for a shorter time. An analogous prescription is used if the individual is in the defection state.

We run the DMM for a time $t$ and we evaluate the mean field
\begin{equation} \label{meanfield}
x(t)  \equiv \frac{\sum_{i}^N \xi_i}{N},
\end{equation}
where  $\xi_i = 1$ or $\xi_i = -1$, according to whether the $i-th$ individual is in the state $C$  or in the state $D$, respectively.   For values of the control parameter much smaller than the critical values $K_c$, which depends on the network topology, the mean field $x$ fluctuates around the vanishing mean values. For values of $K$ significantly larger than $K_c$ the mean field $x(t)$ fluctuates around either $1$ or $-1$. 
It is important to stress that for values of $K$ in the vicinity of the critical value $K_c$, the fluctuations of the mean field have large intensity and the largest intensity corresponds to the critical value $K_c$. The exact value of $K_c$ depends also on the number of units \cite{fourth} and its evaluation is outside the scope of this paper. Here we limit ourselves to notice that  in the case of a regular two-dimensional network $K_c$,  with $N = 100$ is around $1.5$ . The adoption of an irregular networks with a distribution of links departing from the condition  of an equal number of links for each unit may have the effect of significantly reducing the value of $K_c$. A scale-free distribution of links was found \cite{fifth} to make $K_c$ very close to $K_c =  1$, which corresponds to the ideal case where each unit is linked to all the other $N-1$ units. The PDF of the time intervals between two consecutive re-crossings of the origin in the sub-critical condition $K < K_c$ is exponential. At $K = K_c$ the PDF becomes an IPL with power index $\mu=1.5$. In the supercritical state,  the mean field fluctuates around a non-vanishing mean field.  For  $K \gg K_c$ the PDF of the time intervals between two consecutive re-crossing of this non-vanishing mean field again becomes exponential.   The purpose of Section \label{aging} is give the readers a better understanding of the role crucial events, namely non-Poisson renewal events with power index $\mu$ fitting the condition $1 < \mu < 3$. Here we limit ourselves to to illustrate the connection between criticality and multi-fractal spectrum.

\begin{figure}
\includegraphics[width=0.5\textwidth] {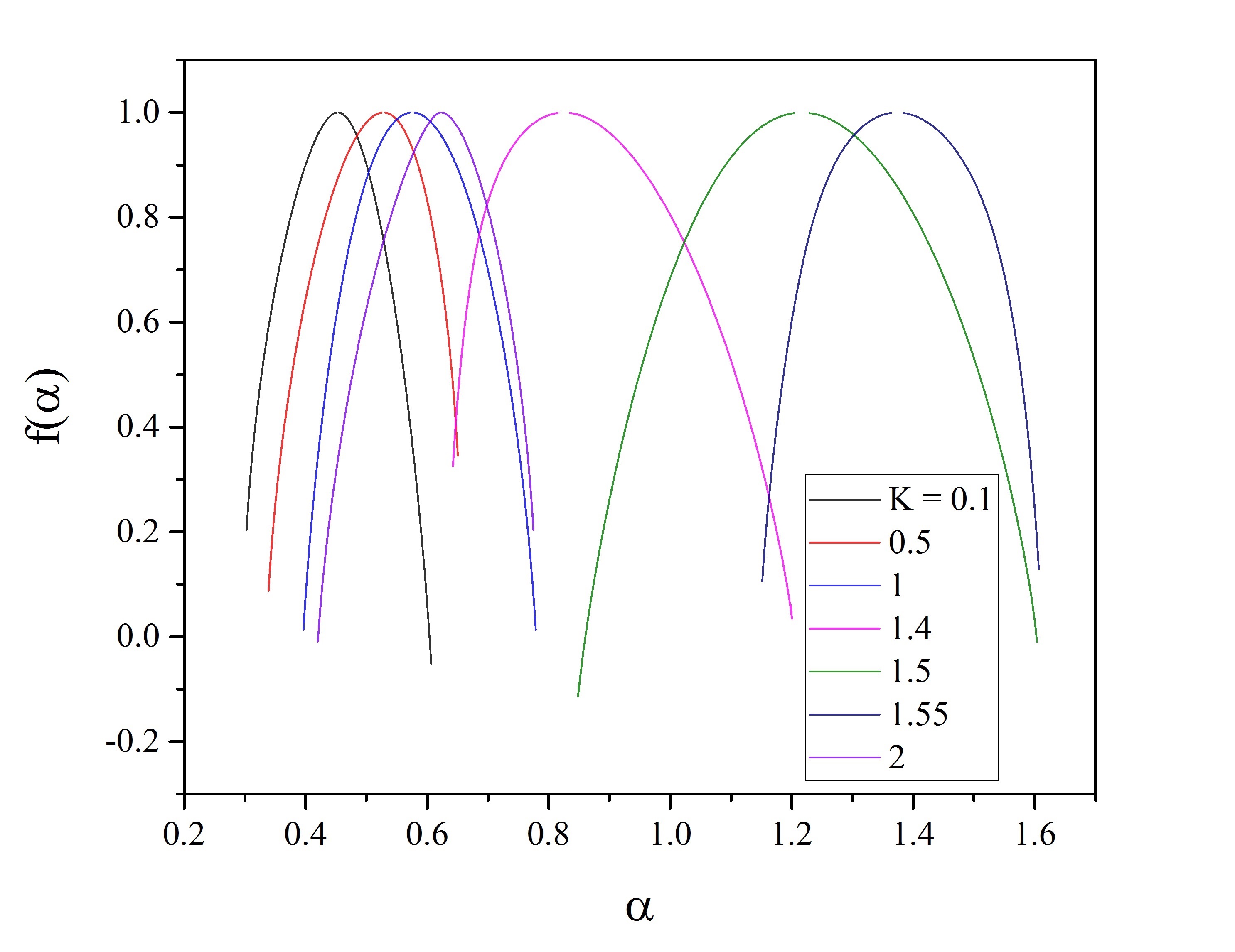}
\caption{Multi-fractal spectrum of the DMM network for different values of the control parameter $K$. Note the non-monotonic behavior of the location of the peak , as well as, the width of the distribution, with the value  of K.}
\label{MultiDMM}
\end{figure}

Following Deligni\`{e}res and co-workers \cite{delignieres} we apply to the time series $x(t)$ the multifractal method of analysis proposed in 2002 by Kantelhardt \emph{et al} \cite{detrended}.  The method of Ref. \cite{detrended} is an extension of the popular technique called Detrended Fluctuation Analysis (DFA) of Ref. \cite{peng} originally proposed to determine the Hurst coefficient $H$. The results depicted in Fig. \ref{MultiDMM} show that the inverted parabola $f(\alpha)$ becomes broadest at criticality. It is interesting to observe that in the sub-critical regime, where the fluctuation of the mean field
of Eq. (\ref{meanfield}) generates ordinary diffusion with Hurst coefficient $H = 0.5$, the spectrum is much sharper and is centered around $\alpha = 0.5$. Increasing the value of the control parameter has the effect of shifting the barycenter  of the inverted parabola towards larger values of $\alpha$ with no significant effect on the parabola's width. The dependence 
of $f(\alpha)$ on $K$ is dramatically non-linear. In  fact, with $K$ going closer to $K_c$ the barycenter of the inverted parabola jumps
to  the vicinity of $\alpha =1.2$ and the parabola, as stated earlier, reaches its maximal width. Moving towards higher values of $K$, supercritical values, has the effect of further shifting to the right the parabola's barycenter. However, the parabola's width becomes much smaller, in line with earlier arguments about the super-critical condition being less complex than the critical condition. It is impressive that with $K = 2$ the parabola's barycenter jumps back to the left, suggesting that 
for even larger values of $K$ complexity is lost, in a full agreement with \cite{timedelay}.

Unfortunately, at the moment of writing this paper, no quantitative theory exists connecting criticality-induced complexity and the emergence of a multi-fractal spectrum at criticality. The work of \cite{gyanendra} suggests that the spectrum $f(\alpha)$ is made broader by the action of crucial events activated by the criticality of the processes of self-organization \cite{sotc}. The numerical results of this section confirms
this property. In fact,  as remarked earlier, Fig. \ref{MultiDMM} shows that the multi-fractal spectrum becomes broadest at criticality, while it becomes sharper in both the supercritical and sub-critical condition, where  
according to \cite{timedelay} the time interval between two consecutive events has an exponential  PDF.
This result suggests that, as done by Deligni\`{e}res and co-workers \cite{delignieres} it may be convenient to measure CME observing 
the correlation between the multi-fractal spectrum of the perturbed network $S$ and the multi-fractal spectrum of the perturbing network $P$ rather than the cross-correlation between the crucial events of $S$ and the crucial events of $P$. This method, although more closely related to the occurrence of crucial events is made hard by ergodicity breaking \cite{piccinini16}, even when the crucial events are visible,  not to speak about the fact that usually crucial events are hidden in a cloud of non-crucial events \cite{heartbeat}.

\subsection{Transfer of multi-fractality from one DMM to another DMM network}
\begin{figure}
\includegraphics[width=0.5\textwidth] {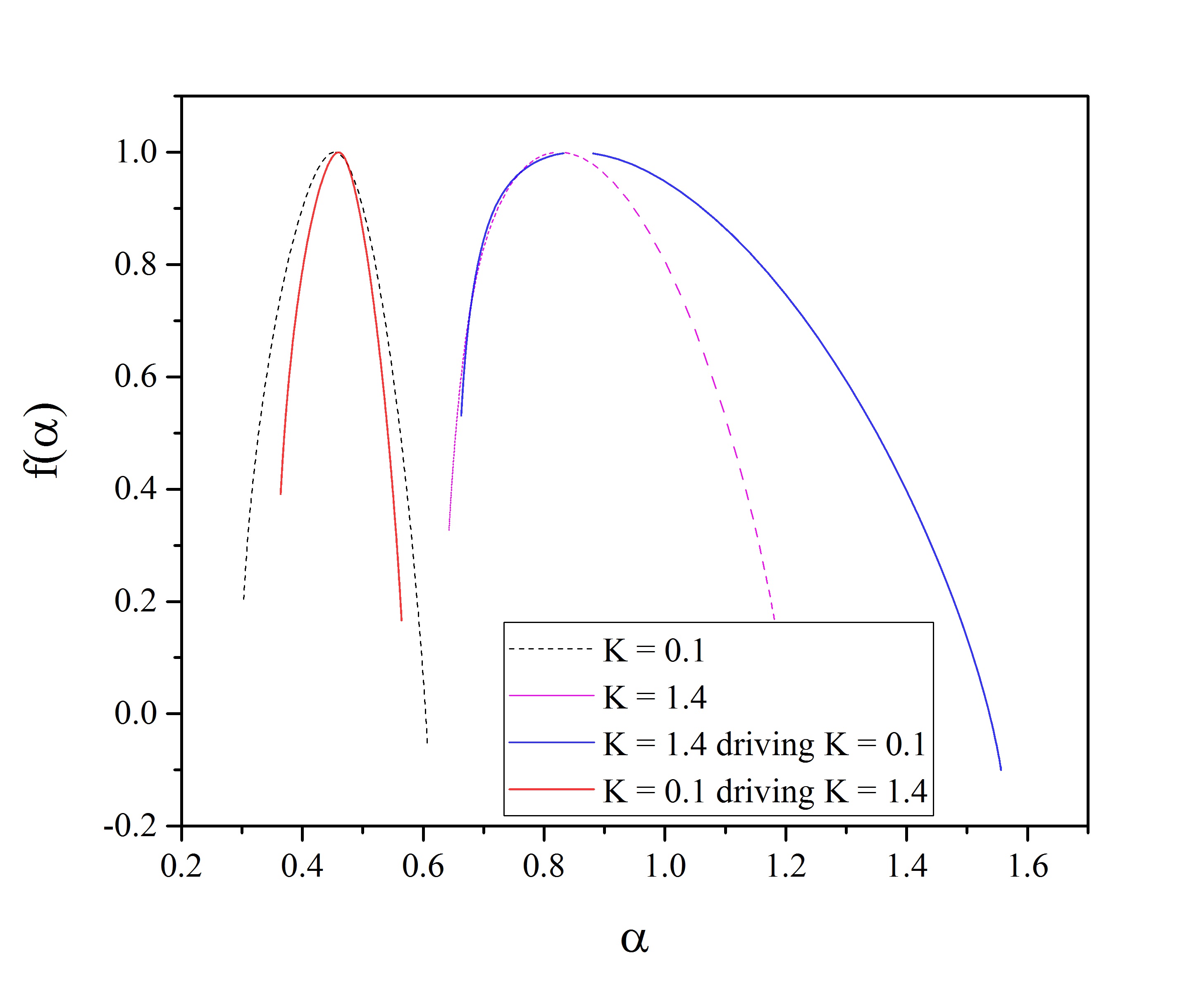}
\caption{The dashed black curve is the multifractal spectrum of the system $A$ with $K = 0.1$ and the dashed pink curve denotes the multifractal spectrum of the system $B$, with $K = 1.4$. One network perturbs the other, as described in the text, with $5\%$ of its units adopting the state $C$ or $D$ according to whether the perturbing system has a positive or a negative mean field. The blue curve is the 
multif-fractal spectrum of the network $B$ under the influence of network $A$ and the red curve is the multiffractal spectrum of $A$ under the influence of $B$. }
\label{DMMMatch}
\end{figure}

In this section we want to discuss an experiment similar to that of  Ref. \cite{delignieres} shown in Fig. \ref{bruce.jpg}. We do that with two DMM networks, the perturbing network playing the role of metronome and the perturbed network playing the role of participants. The complex network $A$ has the control parameter $K = 0.1$, namely, it is a system in the subcritical condition and the network $B$ has the control parameter $K = 1.4$, close to criticality.  We explore two opposite conditions. In the first the system $A$ perturbs the system $B$ and in the second the system $B$ perturbs the system $A$. The perturbation is done as follows. $5\%$ of the units of the perturbed system adopt either the state $C$ or $D$, according to whether the perturbing system has a positive or a negative mean field. 
We see that when the system $B$ with broader spectrum perturbs $A$, with a a sharper spectrum, it forces $A$ to get a much broader spectrum ,even broader the spectrum of $B$. When 
$A$, with a sharper spectrum than $B$ perturbs $B$ it has the effect of making $B$  adopt a spectrum as sharp as that of the perturbing system.  This result can be compared to that of the earlier work of Ref. \cite{second}, where $2\%$ of the units of the perturbed network adopted the choice made by the perturbing network. In that case no correlation was detected between $S$ and $P$ but in the case where both networks being at criticality. This suggests that the correlation between the multi-fractal spectrum $f(\alpha)$ of $S$ and the multifractal spectrum $f(\alpha)$ of $P$ may be a more proper way to study CME \cite{levyflight}.

\section{Detecting Renewal Events\label{aging}}

We devote this Section to establish the equivalence 
between the dynamics of metronome and those of a DMM network  at criticality.  We make a suitable choice of the parameters $\beta$ and $\tau_m$ of Eq. (\ref{metronome}) so as to make it possible
to establish this statistical equivalence. 

\subsection{Renewal character of re-crossings}

As done in earlier work \cite{temporalcomplexity,synchronization} attention
is focused on events corresponding to zero-crossings, that is, to  the time
intervals between succesive crossings of $x=0$. Successive zero-crossings
are used to generate a first time (FT) series $\left\{ \tau _{i}\right\} $,
where $\tau _{i}=t_{i+1}-t_{i}$ is the time interval between two consecutive
events, that is, zero-crossings. An important question about this FT series
is whether a non-zero two-time correlation, between different events,
exists or not. The events are identified as renewal if all two-time and
higher-order correlations are zero.

The renewal nature of the events generated using the metronome equation is
determined by using the aging experiment \cite{aging}. The method,
originally proposed by Allegrini \textit{et al}. \cite{aging}, was for the
purpose of proving that each zero-crossing of a time series is an isolated
event, with no correlation with earlier events. The lack of correlation
implies that when an event occurs, the occurrence of the next event is
completely unrelated and unpredictable; the occurrence of an event can be interpreted as a
form of rejuvenation of the system. Ergodicity breaking \cite{chemphyschem}
is closely connected to the occurrence of renewal events, as can be
intuitively understood by assigning to the waiting-time PDF $\psi (\tau )$
the IPL form 
\begin{equation}
\psi (\tau )\propto 1/\tau ^{\mu }.  \label{PDF}
\end{equation}
In the case $\mu <2$ the mean waiting time is infinite, 
\begin{equation*}
\left\langle \tau \right\rangle =\overset{\infty }{\underset{0}{\int }}\tau
\psi (\tau )d\tau =\infty ,
\end{equation*}
and the longer the total length of the time series under study, the greater
the maximum value of the waiting-time $\tau $ detected. In fact, although a
very short time interval can be drawn immediately after a very long one is
drawn, due to the renewal nature of the process, it is impossible that the
largest value of $\tau $ found, within a sequence of length $L_{1},$ remains
the maximum in examining a sequence of length $L_{2}>L_{1}$. This would
conflict with the $\left\langle \tau \right\rangle =\infty $ condition.

To assess whether the FT series $\left\{ \tau _{i}\right\} $ generated by
Eq. (\ref{metronome}) is detected to be renewal, we generate a second,
auxiliary, time series by shuffling the FT series. We refer to the latter as
the shuffled time series.  We apply the aging experiment algorithm to both the original and the shuffled sequence: We adopt a window of size $t_{a}$, corresponding
to the age of the network that we want to examine. Locate the left end of
the window at the time of occurrence of an event, record the time interval
between the right end of the window and the occurrence of the first event,
emerging after the end of  the window. Note that adopting windows of vanishing size
corresponds to generating ordinary histograms. The histograms generated by 
$t_{a}$ produce different decision-time distribution densities, and these
distribution densities, properly normalized, generate survival
probabilities, whose relaxation can be distinctly different from that of the
ordinary survival probability. 

A non-ergodic renewal process is expected to generate a relaxation that
becomes slower and slower as $t_{a}$ increases. This lengthening of the
relaxation time occurs because the method leads to a truncated time series.
However, the truncation affects the short time intervals more than it does
the long time intervals, thereby reducing the weight of $\psi (\tau )$ for
short times, while enhancing the weight of long time intervals. We have, of
course, to take into account that we adopt normalized histograms. A process
is renewal if the aging of the non-shuffled FT series is identical to the
aging of the shuffled time series.

\begin{figure}
\includegraphics[width=0.5\textwidth]{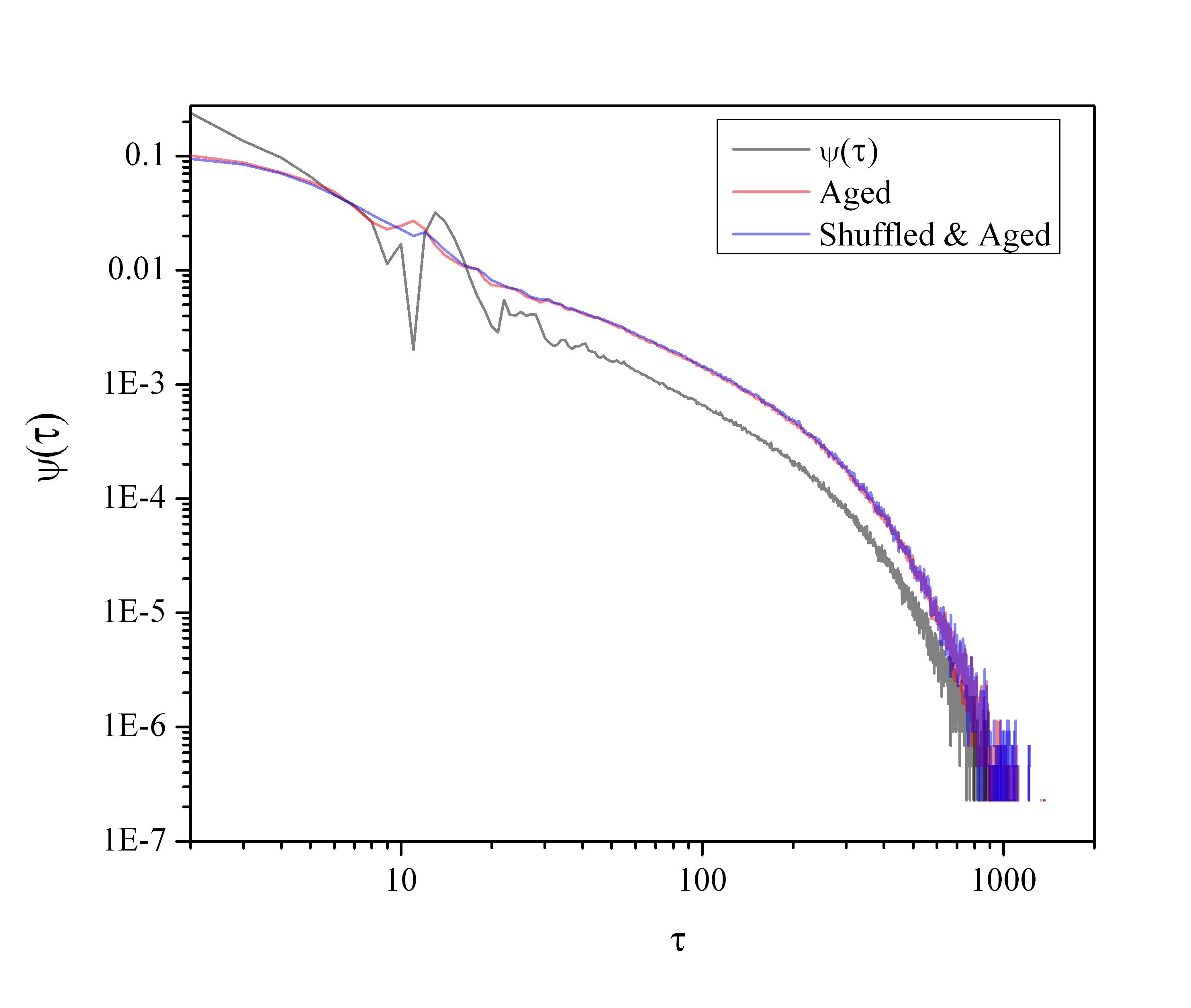}
\caption{(Color online) Waiting-time PDF for $\gamma=1$,  $\beta=200$, $t_m = 10$, $L = 10^7$ and $\tau_a=10.$}
\label{fig_age1}
\end{figure}

\begin{figure}
\includegraphics[width=0.5\textwidth]{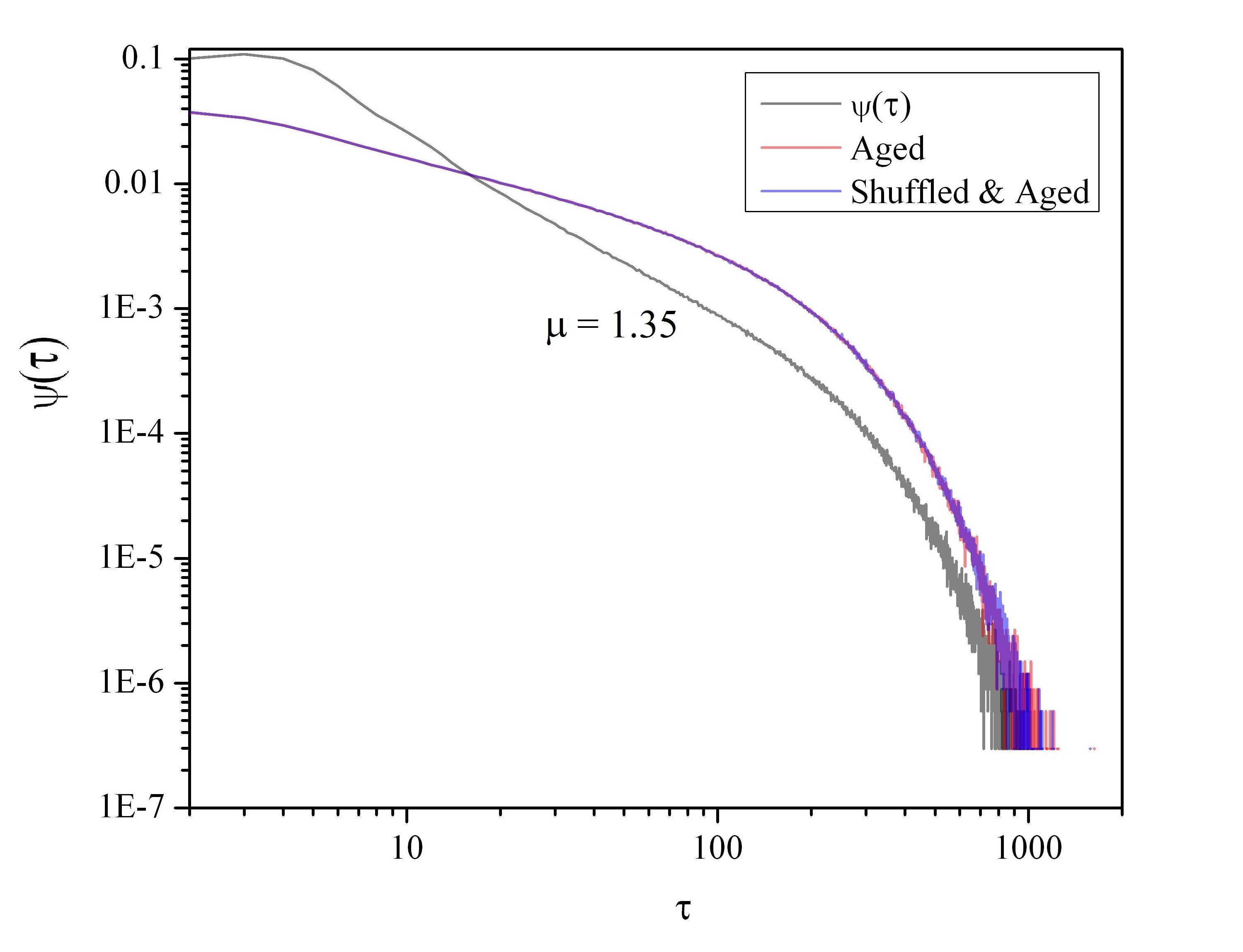}
\caption{Waiting-time PDF for $\gamma=1$, $\beta=1000$,  $\tau_m=1000
$, $L = 10^7$ and $\tau_a=100.$}
\label{fig_age2}
\end{figure}

Let us discuss the results of the aging experiment applied to the time
series $\left\{ \tau _{i}\right\} $ generated by the multi-fractal metronome
data of Eq. (\ref{metronome}). In Fig. \ref{fig_age1} the shuffled time
series is seen to yield a slight deviation from the non-shuffled time series
of approximate magnitude $\tau \sim 10$. This is a consequence of setting
the delay time to $\tau _{m}=10$, thereby establishing a periodicity
interfering with temporal complexity. On the other hand, Fig. \ref{fig_age2}
shows that the shuffled data curve virtually coincides with the non-shuffled
data curve throughout the entire time region explored by the multi-fractal
metronome.

\subsection{Long-time ergodic behavior\label{ergodic}}

Note that the data curves in both Fig.\ref{fig_age1} and Fig.\ref{fig_age2}
are characterized by long-time exponential truncations. In the intermediate
time regions, both conditions show an IPL behavior with IPL-index $\mu =1.5$%
. This property reinforces the conviction that the metronome dynamics and
that of a network of interacting units at criticality are equivalent. In
fact, in the absence of exponential truncation, the complex network would
generate perennial non-ergodic behavior. The latter behavior would make the
network dynamics incompatible with multi-fractality, which, as pointed out by
Juzba and Korbel \cite{challenge}, requires a condition of thermodynamic
equilibrium.

The recent work of Beig et al. \cite{beig} shows that the mean field of a
complex network at criticality generates a mean field $x(t)$, which is well
described by the nonlinear Langevin equation: 
\begin{equation}  
\dot{x}(t)=-a x(t)^{3}+f_{x}(t),  \label{over}
\end{equation}
with $f_{x}(t)$ being a random noise generated by the finite size of the
network, whose intensity is proportional to $1/\sqrt{N}$, and $N$ is the
number of the interacting units. Eq.(\ref{over}) describes the over-damped
motion of a particle within a quartic potential, which has a canonical
equilibrium distribution. However, a particle moving from the initial
condition $x(0)=0,$ undergoes a virtual diffusional process for an extended
time $T_{eq}$. The order of magnitude of this time is given by 
\begin{equation}
T_{eq}\propto   \sqrt{\frac{N}{a}}
\end{equation}
For times $t\ll T_{eq}$ the network's dynamics are non-ergodic, but they
become ergodic asymptotically for $t\gg T_{eq}$.

The zero-crossings are well described by a waiting-time PDF given by Eq.(\ref{PDF}) for times $\tau \ll T_{eq}$. This PDF, however, is exponentially
truncated, and as a consequence of this truncation the renewal aging is not
perennial. As a result of aging the IPL index of the PDF changes from $\mu$ to $\mu-1$, making the decay slower in the intermediate time region, as shown in Fig. \ref{fig_age2}. However, the aging process has also the effect of extending the exponential truncation. In the case of the complex network at criticality, this corresponds to the existence of a thermodynamical equilibrium emerging from the adoption of a large time scale of the order of $T_{eq}$.  It is known  \cite{beig}  that the normalized auto-correlation function of the mean field $x$, to
a high degree of approximation, becomes 
\begin{equation} \label{equilibriumcorrelation}
A (\tau )=exp\left( -\Gamma |\tau| \right) 
\end{equation}
with the relaxation rate given by \cite{beig} 
\begin{equation}
\Gamma \propto a \left\langle x^{2}\right\rangle _{eq}.
\end{equation}
Note that the theory of Ref. \cite{beig} proves that this auto-correlation function 
is obtained by replacing the non-linear Langevin equation of Eq. (\ref{over}) with the following linear Langevin equation
\begin{equation} \label{linear}
\dot x(t) = - \Gamma x(t) + f_{x}(t). 
\end{equation}

In Section \ref{metro} we shall use these arguments to prove that the strong-anticipation of the multi-fractal
metronome may be closely related to the complexity matching observed by
stimulating a complex network at criticality, with the mean field of another
complex network at criticality, see Luko\'{v}ic \emph{et al} \cite{timedelay}
for more details. The authors of the latter article studied a network of $N$
units, with a small fraction of these units, called lookout birds because of
the context of the discussion, or more generally labeled perceiving units, are
sensitive to the mean field of another network of $N$ units in a comparable
physical condition. The units in the both networks are decision making
individuals, who have to make a dyadic choice, between the yes, $+1$, and
no, $-1$, state. The perceiving units adopt the $+1$ state, if the mean field
perceived by them is positive, $y>0$, or the $-1$ state, if they perceive $y<0$. The cross-correlation between the mean field $x(t)$ of the driven
network and the mean field $y(t)$ of the driving network attains maximal
intensity when both networks are at criticality. It turns out that the
cross-correlation function is identical to the auto-correlation function of 
$x(t)$, with a significant shift, namely,  the cross-correlation function  $\left\langle
x(t+ \tau)y(t )\right\rangle $ gets its maximal value when $\tau = \Delta$, where the delay $\Delta$ represents a delay in transmitting information from the perturbing to the perturbed complex
network. 

An intuitive interpretation of the above time delay is that the information
perceived by the lookout birds must be transmitted to all the units of their
network. Luko\'{v}ic \emph{et al} \cite{timedelay} adopted a different
interpretation of this important delay time. To vindicate their view they
studied the all-to-all coupling condition, where the preceptors are coupled
to all the other units in their network. Even in this case the
cross-correlation function is characterized by a significant time delay,
with respect to the unperturbed correlation function of $x(t)$. The reason
for the delay is that the group, in the case discussed by \cite{timedelay,cognition},
a flock of birds, can follow the direction of the
driving network only when a significantly large number of zero-crossings
occur. A zero-crossing corresponds to a free-will condition, where the whole
network, can be nudged by an infinitesimally small fluctuation, to select
either the positive or the negative state. Thus, the renewal nature of the
zero-crossing events becomes essential for the emergence of such cooperative
behavior as cognition \cite{cognition}, as in changing one's mind for no
apparent reason. The zero-crossing is the time at which the network is most sensitive to a perturbation, so the more zero-crossings the shorter the interval between a perturbation and response.

\subsection{Beyond ordinary diffusion}

We note that the waiting-time PDF of Fig. \ref{fig_age2} is given by
\begin{equation}
\mu \approx 1.35
\end{equation}
This does not conflict with the earlier arguments on the exponential nature of the infinitely aged regime. In fact, on the basis of a theoretical approach based on the observation of random growth of surfaces \cite{failla} it is argued \cite{rohisha} that in all the organization processes  the fluctuation of $x(t)$ around the origin are non-Poisson renewal events characterized by the following common property.  Let us call $\Psi(t)$ the probability that 
no renewal non-Poisson event occurs at a distance $t$ from an earlier event. The Laplace transform of $\Psi(t)$, $\hat \Psi(u)$, is given by
\begin{equation} \label{rohisha}
\hat \Psi(u) = \frac{1}{u + \lambda^{\alpha} \left(u + \Delta\right)^{1-\alpha}}.
\end{equation}

On the basis of Fig. \ref{fig_age1}  we assume that there exists a wide time interval generating the PDF index $\mu = 1 + \alpha$, which is $1.35$ in the case of that figure.
This time interval is defined by
\begin{equation} \label{rohisha3}
\frac{1}{\lambda} \ll t \ll \frac{1}{\Delta}.
\end{equation}
This wide time interval in the Laplace domain becomes
\begin{equation}
\lambda  \gg  u  \gg \Delta,
\end{equation}
thereby turning Eq. (\ref{rohisha}) into
\begin{equation} \label{rohisha2}
\hat \Psi(u) = \frac{1}{u + \lambda^{\alpha} u^{1-\alpha}}.
\end{equation}

This is equivalent to the Laplace transform of the Mittag-Leffler function, which is known to be a stretched exponential in the time regime $t < 1/\lambda$ and the inverse power law 
$1/t^{\alpha}$ in the time regime $t > 1/\lambda$. Due to Eq. (\ref{rohisha3}) we conclude that in that time interval 
the waiting time distribution density is an inverse power law with $\mu = 1 + \alpha$ while for times $t > 1/\Delta$ 
\begin{equation} \label{theor2}
\psi(t) = \Gamma exp\left(- \Gamma t\right),
\end{equation}
where
\begin{equation} \label{Gammanow}
\Gamma \equiv \lambda^{\alpha} \Delta ^{1-\alpha}. 
\end{equation}
Note that the corresponding survival probability, $\Psi(t)$, 
gets the form 
\begin{equation} \label{theory}
\Psi(t) =  exp\left(- \Gamma t\right)
\end{equation}
and is identical an equilibrium correlation function with the same form as that of Eq. (\ref{equilibriumcorrelation}),  with $\Gamma$ given by Eq. (\ref{Gammanow}).

In the example discussed in Section \ref{metro}, where $\Gamma = 0.01$, we have $\Delta = 0.0078$ corresponding to the time $t = 129$.

\begin{figure}
\includegraphics[width=0.5\textwidth] {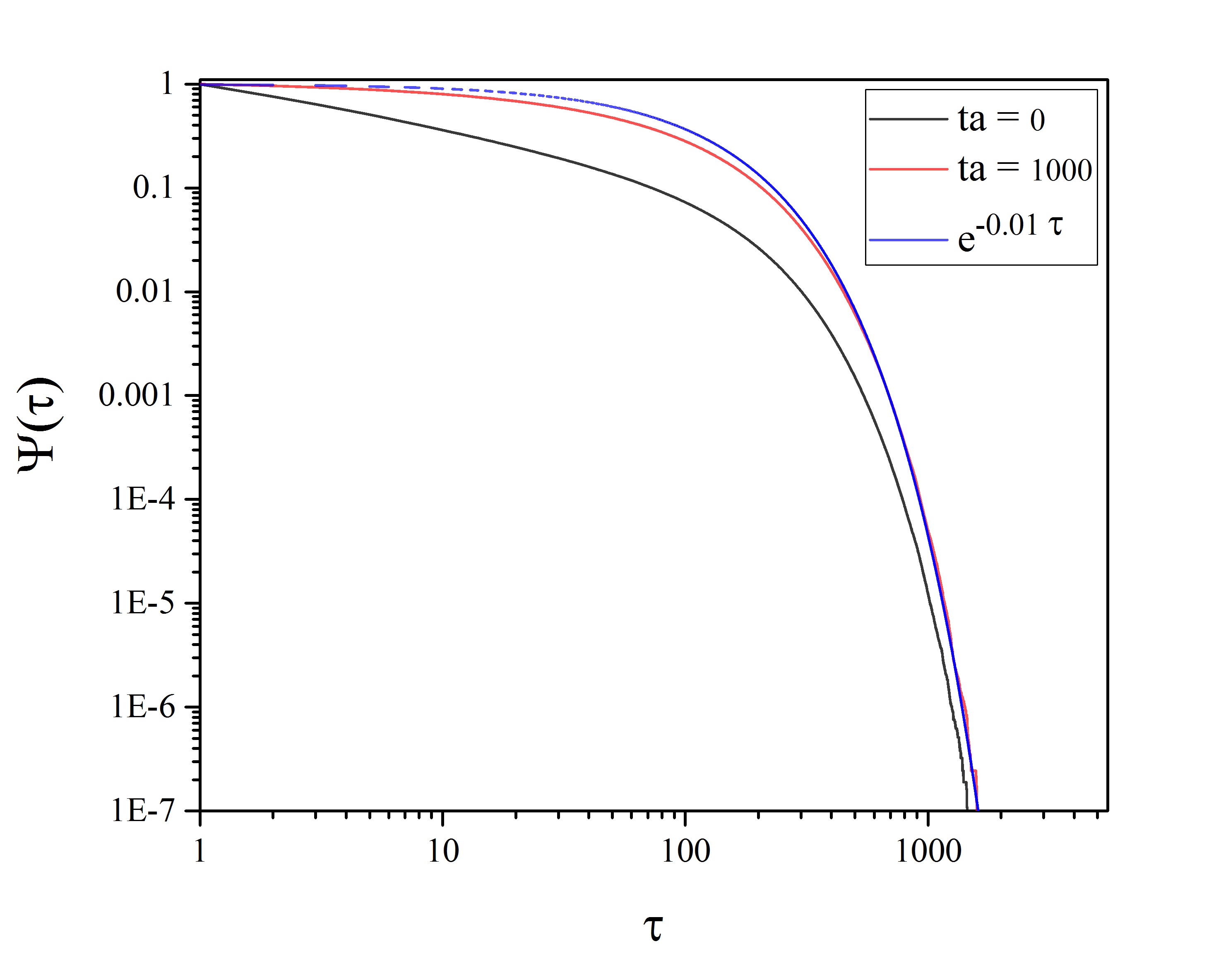}
\caption{Survival probability of the crucial events of the metronome, compared to the exponential function corresponding to infinitely large age.The blue curve is the exponential function of Eq. (\ref{theory}).}
\label{Survival}
\end{figure}

In Fig. \ref{Survival} we make a comparison between the numerical results on the aging of the renewal events and the theoretical prediction of Eq. (\ref{theory}). 
The good agreement between numerical results and numerical prediction confirms that the metronome hosts crucial events therefore supporting our conviction that
the multi-fractal properties of the metronome, stressed by the work of Deligni\'{e}res and coworkers \cite{delignieres}, are a manifestation of the action of crucial events.

We afford a further support to this important property studying the spectrum $S(\omega)$ of the metronome in the condition corresponding to Fig. \ref{fig_age2}.  
According to the arguments illustrated in Section \ref{spectrum}, see Eq. (\ref{mirkospectrum}), the $1/f$-noise generated by the metronome in this condition, with $\mu = 1.333$,
should yield $\nu = 1.67$.  
\begin{figure}
\includegraphics[width=0.5\textwidth] {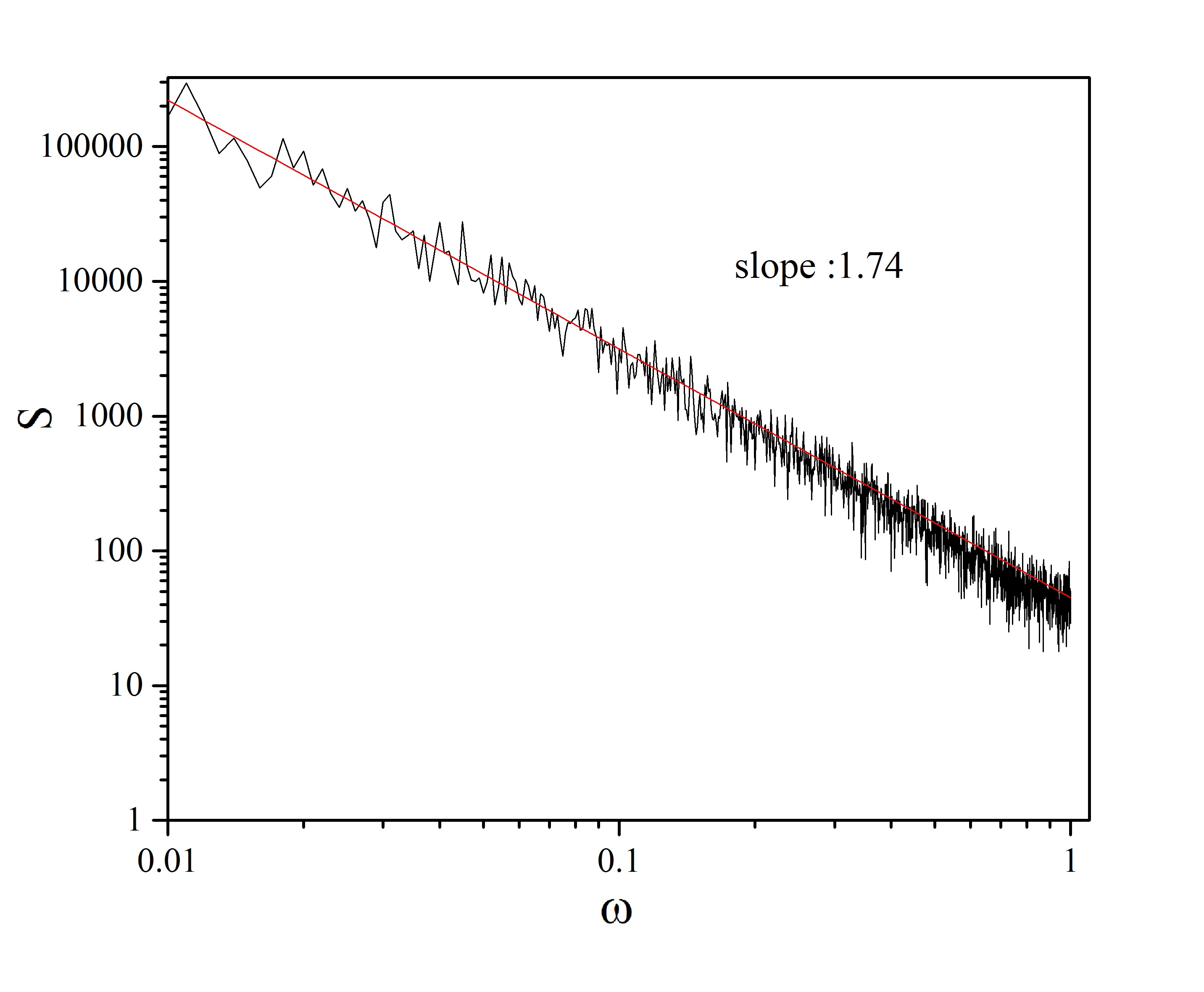}
\caption{The spectrum $S(\omega)$ of the metronome in the condition  of Fig. \ref{fig_age2}.}
\label{spectrum}
\end{figure}
Fig. \ref{spectrum} yields a satisfactorily agreement between theory and numerical results if we take into account the challenging numerical issue of evaluating the spectrum of a non-ergodic process
with the density of crucial events decreasing upon increase of $L$.

\section{Complexity Matching between two multi-fractal metronomes} \label{metro}

In this Section we discuss the results of a numerical experiment done on the cross-correlation between two identical multi-fractal metronomes,
in the condition illustrated by Fig. 2, which makes them, as shown in the earlier Section, equivalent to the complex networks at criticality of Ref. \cite{beig}. 
We stress that this equivalence rests on sharing the same temporal complexity, namely, the same non-Poisson renewal statistics for the zero-crossings. 
The results of this experiment of information transport from one to another identical multi-fractal metronome generates a qualitative agreement 
with the results of the earlier work of Ref. \cite{timedelay} on the information transport from a complex network at criticality to another complex network at criticality.
However, we go much beyond this qualitative agreement. To do so, we use the theory of Ref. \cite{beig} to derive an analytical expression for the cross-correlation used to evaluate the information transport and compare it to the cross-correlation
between the two equivalent multi-fractal metronomes done in this section, and we find outstanding agreement.

The numerical calculations of this section are based on the following set of coupled equations:

\begin{equation}  
\dot{x}=-\gamma x(t)-\beta sin\left( x(t-\tau _{m})\right) +\chi y,
\label{11}
\end{equation}
where $x(t)$ is the mean field of the responding network and $y(t)$ is the
mean field generated by the driving network

\begin{equation}
\dot{y}=-\gamma y(t)-\beta sin\left( y(t-\tau _{m})\right).  \label{12}
\end{equation}
This choice of equations is done to mimic the influence that perceptors
(lookout birds) exert on their own network in response to an external
network. The interaction term $\chi y$ must be weak to mimic the influence
of a very small number of perceiving units. For this reason we assign to the
coupling coefficient $\chi $ the value $\chi =0.1$.

 \begin{figure}
\includegraphics[width=0.5\textwidth]{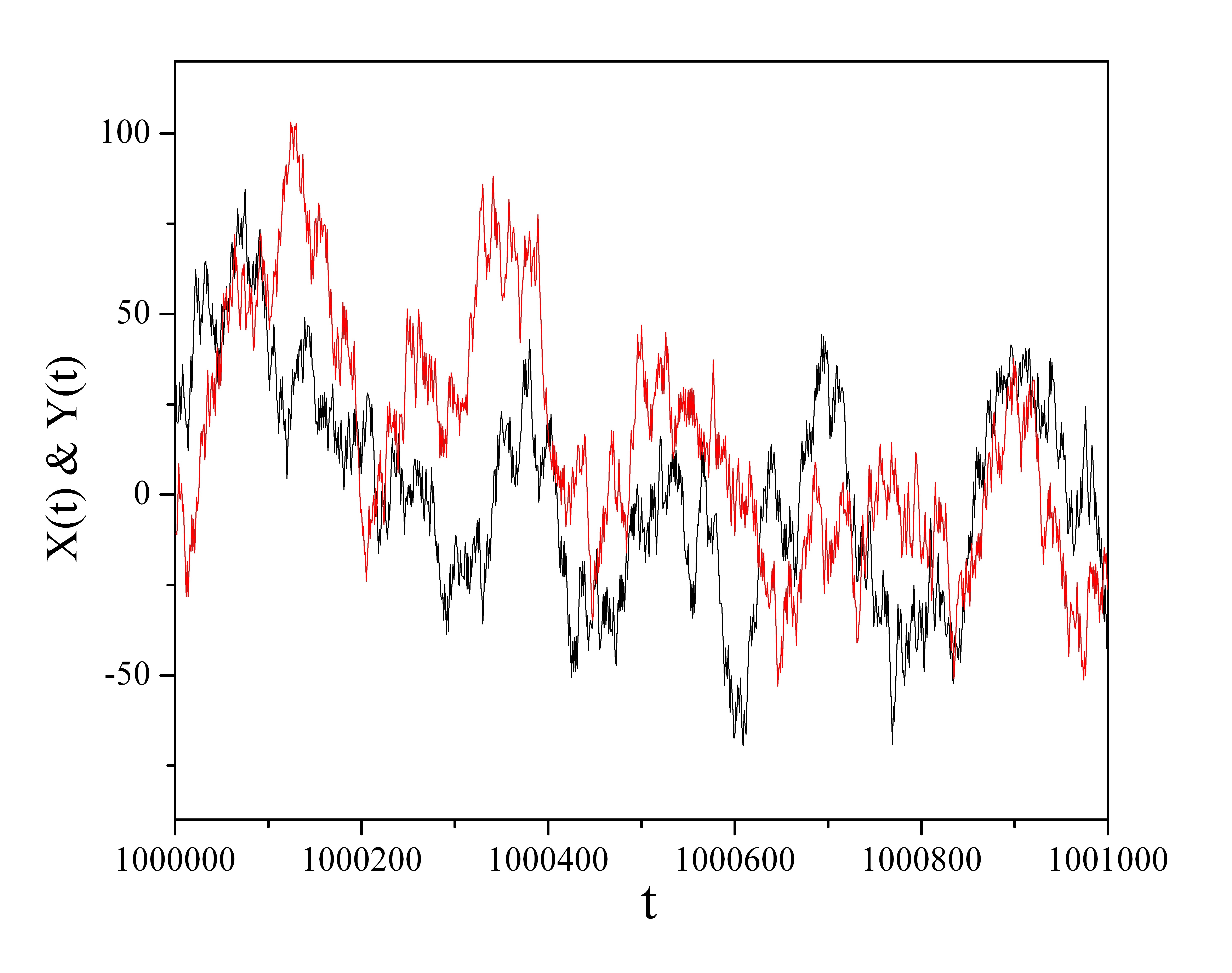}
\caption{(Color online) Complexity matching between driven and driving
metronome. The black curve is the driving system and red curve corresponds to the driven network.}
\label{fig_synchronization}
\end{figure}

In Fig. \ref{fig_synchronization} we show that this weak coupling results in a remarkable synchronization of the driven metronome with the driving metronome. Let us now move a quantitative discussion. 
To make our results compatible with the observation of single complex systems, the brain being an example of unique system making it impossible for us to adopt the ensemble average method, we use the time average approach and we define the autocorrelation function $A(\tau)$ and the cross-correlation function $C(\tau)$ as follows:
\begin{equation} \label{auto}
A(\tau) = \frac{\int_{0}^{T-\tau} dt x(t +\tau) x(t)}{T-\tau}
\end{equation}
and
\begin{equation} \label{cross}
C(\tau) = \frac{\int_{0}^{T-\tau} dt x(t + \tau) y(t)}{T-\tau}.
\end{equation}
In both cases we set $T = 10^7$. The auto-correlation function of Eq. (\ref{auto}) is evaluated with $\chi=0$, namely, when the metronome $x$ is not perturbed by the metronome $y$.

\begin{figure}
\includegraphics[width=0.5\textwidth]{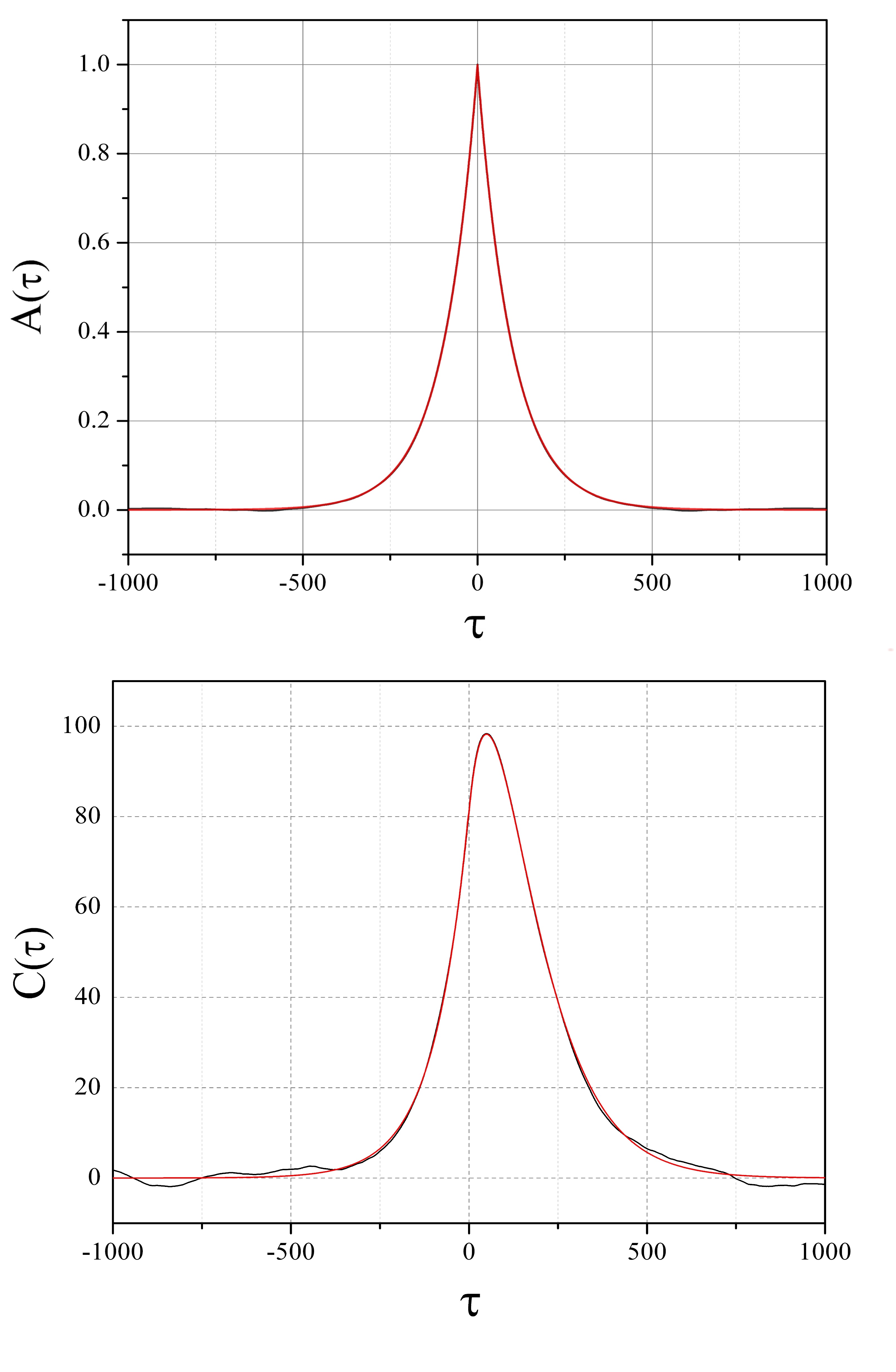}
\caption{(Color online) Top panel: Numerical  auto-correlation function $A(\tau)$ of Eq. (\ref{auto}). The numerical value  of $\Gamma$ is $\Gamma = 0.01$. Bottom panel: The black curve denotes the numerical result for the cross-correlation function with  $\gamma=1$,
$\beta=1000$ and $\tau_m=1000$. The red curve is derived from Eq. (\ref{earlier1}) and Eq. (\ref{earlier2}) with the same $\Gamma$ as in the top panel and the fitting parameter $b$ with the value $b = 81.8$. }
\label{fig_negativetail}
\end{figure}

 The bottom panel of Fig. \ref{fig_negativetail}  shows that
the cross-correlation function, as expected, is characterized by a
significant delay, on the order of $\tau \sim 100$. The cross-correlation
function is asymmetric with respect to the shifted maximum. Notice that we have selected the value $\tau _{m}=1000$, of Fig. \ref{fig_age2}   so as to reduce the
influence of periodicity on the temporal complexity.

Let us now generate an analytical expression to match these numerical results. First of all let us stress that setting $T = 10^7$ is equivalent to make the numerical observation in the ergodic regime, where time and ensemble averages are expected to yield the same results. The adoption of ensemble averages make the calculations much simpler and for this reason, with no contradiction with the statement that 
we focus our attention on unique complex networks, we rest our theoretical arguments on  ensemble averages. 

The assumption that the time series generated by the multi-fractal metronome
and that of the complex network are equivalent at criticality, and the arguments \cite{beig}  proving the equivalence between Eq. (\ref{over}) and Eq. (\ref{linear})   lead us 
 to replace Eq. (\ref{11}) and Eq. (\ref{12}) with the linearized forms

\begin{equation}
\dot{x}=-\Gamma x(t)+f_{x}(t)+\chi y,  \label{21}
\end{equation}%
and 
\begin{equation}
\dot{y}=-\Gamma y(t)+f_{y}(t),  \label{22}
\end{equation}%
where $f_{x}(t)$ and $f_{y}(t)$ are mutually uncorrelated Wiener noises. It
is straightforward to show, using the lack of correlation between the two
sources of noise, that the stationary cross-correlation function $C(\tau)$ is 
\begin{equation}
C(\tau )\equiv \lim_{t \rightarrow  \infty} \left\langle x(t+\tau )y(t)\right\rangle = \lim_{t \rightarrow  \infty}\chi
\int_{0}^{t+|\tau |}dt^{\prime} e^{ -\Gamma (t+|\tau |-t^{\prime })}\left\langle
y(t)y(t^{\prime })\right\rangle .  \label{correlation}
\end{equation}%
In the absence of coupling, the two metronomes are characterized by the
normalized auto-correlation functions 
\begin{equation}
A_{x}(\tau)= A _{y}(\tau)= e^{-\Gamma |\tau |}
\equiv A(\tau).  \label{nest}
\end{equation}
As a consequence 
\begin{equation}
\left\langle y(t)y(t^{\prime })\right\rangle =\left\langle
y^{2}\right\rangle _{eq}e^{-\Gamma |t-t^{\prime }|}.  \label{toplug}
\end{equation}
By inserting Eq. (\ref{toplug}) into Eq. (\ref{correlation}), and taking
into account that for $\tau >0$, there are two distinct conditions, 
$t^{\prime }<t$ and $t<t^{\prime }<t+\tau $, we obtain 
\begin{equation}  
C(\tau )\equiv \lim_{t \rightarrow  \infty} \left\langle x(t+\tau )y(t)\right\rangle =be^{-\Gamma |\tau |},\text{  }\tau <0,  \label{earlier1}
\end{equation}
and 
\begin{equation}
C(\tau )\equiv \lim_{t \rightarrow  \infty} \left\langle x(t+\tau )y(t)\right\rangle =b e^{-\Gamma |\tau |}(1 + 2 \Gamma \tau),\text{ }\tau >0.
\label{earlier2}
\end{equation}%
Note that 
\begin{equation}
b\equiv \frac{\left\langle y^{2}\right\rangle _{eq}\chi }{2\Gamma }.
\end{equation}

It is important to stress that the normalized auto-correlation function of the multi-fractal metronome 
\begin{equation} \label{auto}
A(\tau) \equiv \lim_{t \to \infty} \frac{\left<x(t) x(t+\tau)\right>}{\left<x(t)^2\right>} = e^{- \Gamma |\tau|} \end{equation}
is evaluated numerically and it is illustrated in the top panel of Fig. 5.  We derive the value of $\Gamma $ from this numerical treatment and its value $\Gamma = 0.01$ is used in Eq. (\ref{earlier1}) and in Eq. (\ref{earlier2}). As a consequence, to
make a comparison between theoretical and numerical cross-correlation
function we have only one fitting parameter, $b$, the intensity of
autocorrelation function at $\tau =0$. The bottom of Fig. \ref{fig_negativetail} depicts
the comparison between numerical and theoretical results and shows that
the agreement between the two goes far beyond the qualitative.
It is interesting to notice Eq. (\ref{earlier1}) yields for the time shift $\Delta$ of the cross-correlation function the following analytical expression
\begin{equation}
\Delta = \frac{1}{2 \Gamma}.
\end{equation}
This interesting expression shows that reducing $\Gamma$ has the effect of increasing the time shift. On the other hand reducing $\Gamma$ has the effect of making the non-ergodic time regime $t < T_{eq}$ more extended,
thereby suggesting a close connection between complexity matching and ergodicity breaking.

\section{Transfer of multi-fractal spectrum   from a  complex to a deterministic metronome } \label{complexDRIVINGdeterministic}

After illustrating the similarity between the metronome-metronome interaction and the DMM-DMM interaction when both DMM networks are at criticality, let us move to discuss the transfer of information from a complex metronome to a a deterministic metronome. On the basis of the PCM we should expect that in this case no significant transfer of information occurs.  See, for instance, the earlier work of Ref. \cite{second}, as an example of a lack of information transport when the DMM driving network is at criticality and the driven DMM network is in the subcritical condition.

 \begin{figure}
\includegraphics[width=0.5\textwidth]{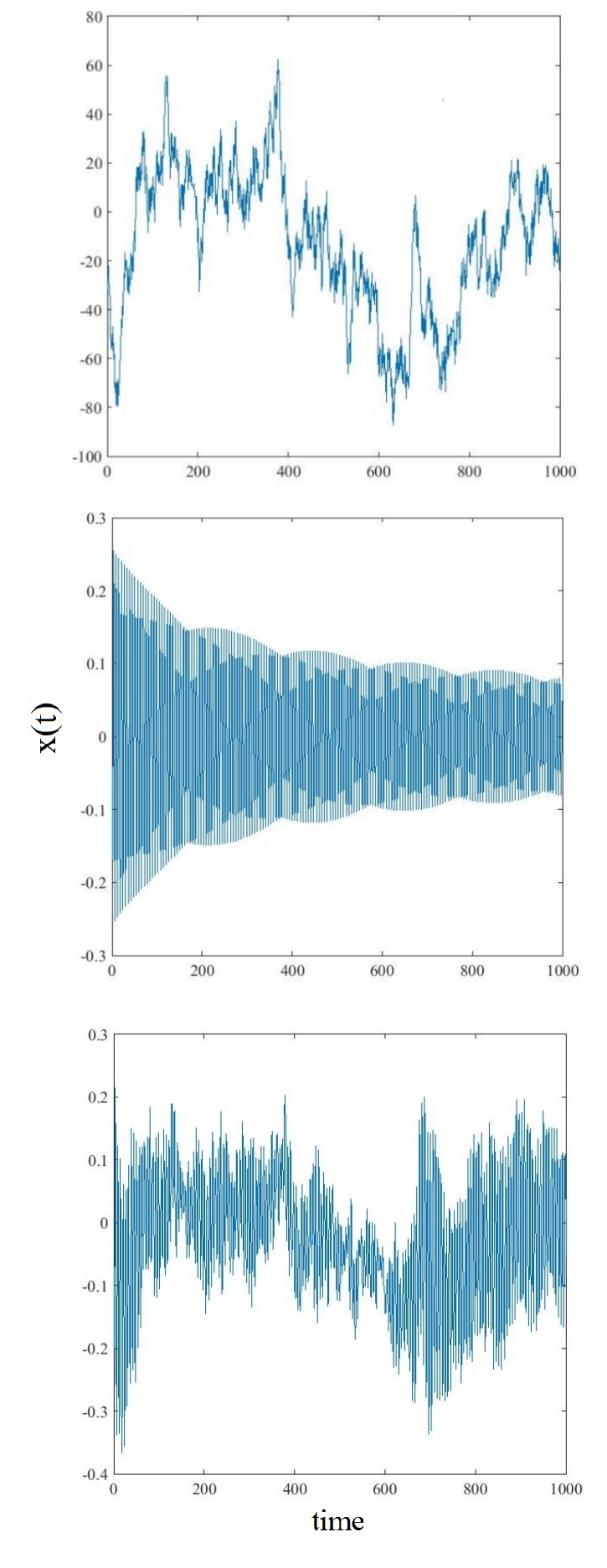}
\caption{(Color online) Top panel: Time evolution of driving metronome with  $\gamma = 1$,  $\beta=1000$ and $\tau_m=1000$.  Middle panel:  Time evolution of driven metronome (before connection) with  $\gamma = 1$,  $\beta=100$ and $\tau_m=1$. Bottom panel: Time evolution of driven metronome (after connection,  $\chi = 0.1$). }
\label{threepanels}
\end{figure}

The field $x(t)$ of the driving network is illustrated by the top panel of Fig.\ref{threepanels}. The waiting time PDF between
two consecutive regressions to the origin is of the same kind as that illustrated in Fig. \ref{fig_age1}, with an intermediate time region with the complexity $\mu = 1.35$,  and an exponential truncation. The driven metronome in the absence of 
the influence of the driving metronome generates the field $x(t)$ illustrated by the middle panel of Fig.\ref{threepanels}. This is a fully deterministic condition corresponding to the choice of $\tau_m = 1$, which implies a lack of complexity. 
More precisely,  Eqs. (\ref{11}) and (\ref{12}) has been changed into

\begin{equation}  
\dot{x}=-\gamma x(t)-\beta sin\left( x(t-\tau _{S})\right) +\chi y,
\label{111}
\end{equation}
and 
\begin{equation}
\dot{y}=-\gamma y(t)-\beta sin\left( y(t-\tau _{P})\right)  \label{121},
\end{equation}
respectively, where $\tau_{S} = 1$ and $\tau_{P} = 1000$.

 \begin{figure}
\includegraphics[width=0.5\textwidth]{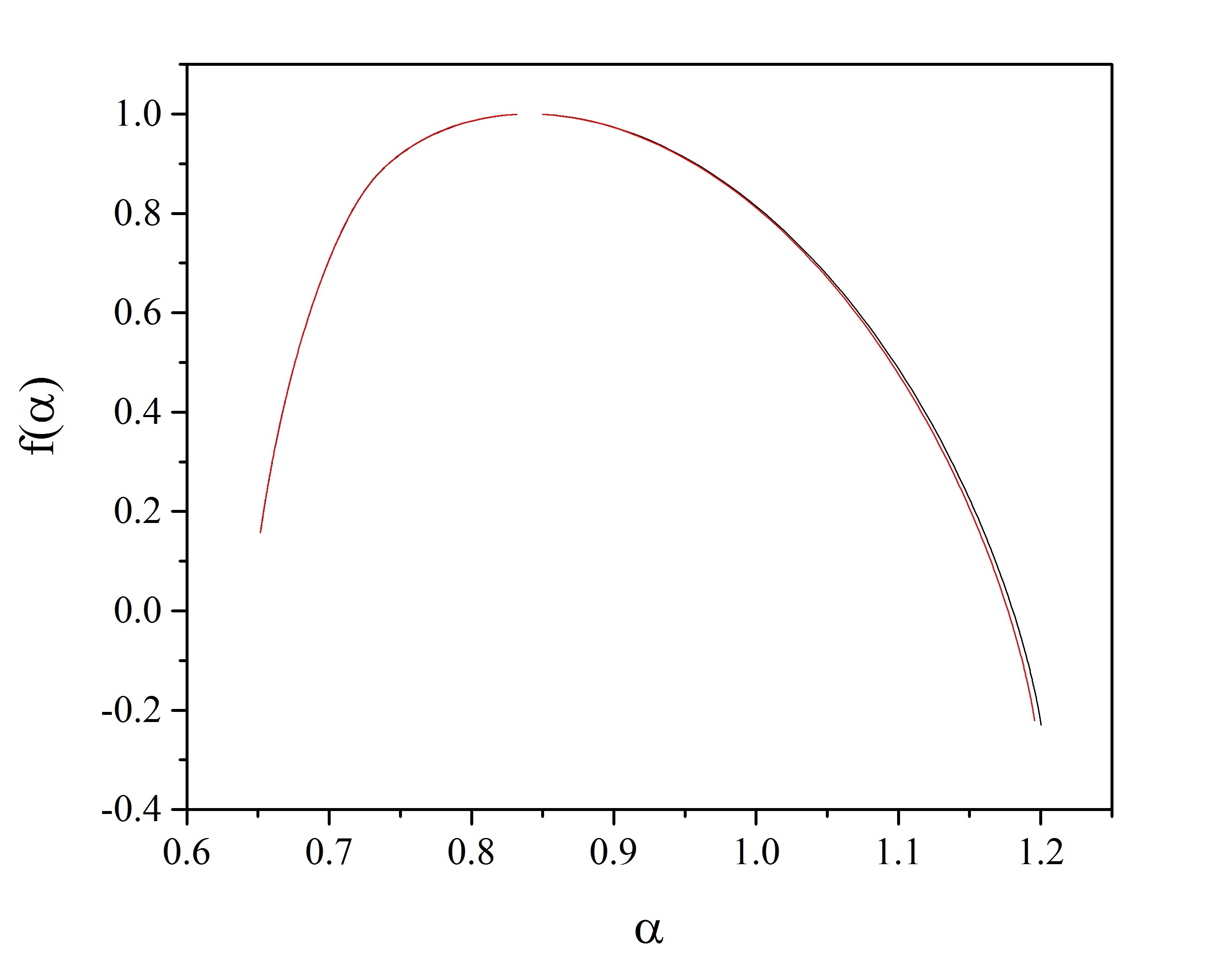}
\caption{(Color online) Black curve: Parabola of driving metronome with $\gamma = 1$,  $\beta=1000$ and $\tau_m=1000$. Red curve:  Parabola of driven metronome with $\gamma = 1$,  $\beta=100$ and $\tau_m=1$. $\chi = 0.1$. }
\label{lastparabola}
\end{figure}

The influence of the driving metronome on the driven one is illustrated by the bottom  panel of Fig.\ref{threepanels}. It is evident that the driven metronome has absorbed the complexity of the driving metronome. This important property is made compelling from the results of Fig. \ref{lastparabola}. This figure was obtained by applying the multi-fractal algorithm of Ref. \cite{detrended} to the time series $x(t)$ of the driven metronome moving under the influence of 
the driving metronome.

How to explain this surprising result?  The earlier dynamical work on complexity matching was based on the assumption that a complex network made complex by criticality is characterized by $K = K_c$ and a network with no complexity has a control parameter distinctly smaller than the critical value $K_c$. The crucial events of the driving network exert an influence on the time occurrence of crucial events of the driven network and consequently a form of synchronization takes place, when both networks are at criticality.  However, the interaction between the two networks does not affect the values of their control parameters. If the driven network is in the subcritical regime, the statistics of its events remain of Poisson kind, thereby making it impossible to realize such a synchronization effect.  The parameter $\tau_m$ apparently plays the same role as the control parameter $K$ of the DMM network, but  $\beta sin\left( x(t-\tau _{S})\right) +\chi y$ of Eq. (\ref{111}), with $\tau_{S} = \tau_m$  under the influence of perturbation is turned into $\beta sin\left( y(t-\tau _{P})\right)$. The multi-fractal metronome is more flexible than a DMM network in the subcritical regime.

\section{Concluding Remarks}\label{Remarks}

The main results of this paper are the following. The multi-fractal metronome used by Deligni\'{e}res and co-workers \cite{delignieres} is equivalent to a complex network at crtiticality in the sense that it hosts crucial events. This establishes a connection between multi-fractal spectrum $f(\alpha)$ and crucial events. This result is in a qualitative accordance with the observation done in Fig. \ref{MultiDMM}. The multi-fractal metronome is equivalent to a complex network at criticality, although its temporal complexity is characterized by $\mu = 1.35$ rather than $\mu = 1.5$, as for DMM \cite{beig}. 
The cross-correlation between two identical networks in the long-time limit is indistinguishable from that of two DMM networks at criticality, as shown by Fig. \ref{fig_negativetail}. 
Thus, we can conclude that the complexity matching established by Deligni\'{e}res and co-workers \cite{delignieres} and illustrated in Fig. \ref{bruce.jpg} is a process made possible by the influence that the crucial events of the metronome exert on the crucial events of the brain of the participants. 

The adoption of $f(\alpha)$ as a measure of the response of $S$ to $P$ seems to be more powerful than the use of cross-correlation function. 
In fact, It also of remarkable interest to notice that the correspondence between crucial events and broader distribution of $f(\alpha)$ makes it possible to establish the existence of a correlation between the perturbed network $S$ and the perturbing network $P$ in conditions far from the complexity matching of both networks at criticality, where earlier work did not reveal any significant correlation \cite{second} .
The metronome in the physical condition making it equivalent to a network of interacting units at criticality exerts a strong influence on a network in the deterministic condition, see Fig. \ref{threepanels}. Also in this case $f(\alpha)$ is a powerful indicator of correlation. Fig. \ref{lastparabola} shows that the perturbed deterministic metronome inherits the spectral distribution $f(\alpha)$ of the perturbing metronome.

{\bf \emph{Acknowledgments}}.
The authors are grateful to David Lambert for contributing important discussions on the subject of this article and for his technical assistance on the preparation of the manuscript. 
KM and PG  warmly  thank ARO  and Welch for support through Grant No. W911NF-15-1-0245 and Grant No. B-1577, respectively.


\begin{references}

\bibitem{west} B. J. West, E. L. Geneston, P. Grigolini, \emph{Maximizing information exchange between complex networks}, Phys. Rep. {\bf 468}, 1 (2008). 

\bibitem{delignieres12} D. Deligni\`{e}res and V. Marmelat, \emph{Fractal
fluctuations and complexity: Current debates and future challenges},
Critical Rev. Biomed. Eng. {\bf 40}, 485 (2012).

\bibitem{aquino10} G. Aquino, M. Bologna, P. Grigolini and B.J. West,
\emph{Beyond the death of linear response: 1/f optimal information transport}, 
Phys. Rev. Lett. \textbf{105, }069901 (2010).

\bibitem{mirko} M. Lukovic,  P. Grigolini, 
\emph{Power spectra for both interrupted and perennial aging processes} J. Chem. Phys, {\bf 129}, 184102  (2008).   


\bibitem{original1} S. Bianco, E. Geneston, P. Grigolinia, M. Ignaccolo, \emph{Renewal aging as emerging property of phase synchronization},  
Physica A {\bf 387}, 1387 (2008).

\bibitem{synchronization}  M. Turalska, M. Lukovic, B. J. West, P. Grigolini, \emph{Complexity and synchronization}, Phys. Rev E {\bf 80}, 021110 (2009). 


\bibitem{beig} M. T. Beig, A. Svenkeson, M. Bologna, B. J. West, P. Grigolini, \emph{Critical slowing down in networks generating temporal complexity}, Phys. Rev.  E {\bf 91}, 012907 (2015). 



\bibitem{aquino11} G. Aquino, M. Bologna, B.J. West and P. Grigolini, \emph{Transmission of information between complex systems: 1/ f resonance},
Phys. Rev. E, \textbf{83}\textit{, }051130 (2011).



\bibitem{piccinini16} N. Piccinini, D. Lambert, B. J. West, M. Bologna and P.
Grigolini, \emph{Nonergodic complexity management}, Phys. Rev. E, \textbf{93}, 062301 (2016).

\bibitem{delignieres}  D. Deligni\`{e}res, Z. M. H. Almurad,  C. Roume, V. Marmelat, \emph{Multifractal signatures of complexity matching}, Experimental Brain Research, {\bf 234}, 2773 (2016).

\bibitem{mandelbrot77} Mandelbrot B.B., \textit{Fractals: Form, Chance and
Dimension}, W.H. Freeman, San Francisco (1977).


\bibitem{scafetta} N. Scafetta, P. Grigolini, \emph{Scaling detection in time series: Diffusion entropy analysis}, Phys. Rev. E {\bf 66}, 036130  (2002). 




















\bibitem{soma03} R. Soma., D. Nozaki, S. Kwak, and Y. Yamamoto, \emph{1/f Noise Outperforms White Noise in Sensitizing Baroreflex Function in the Human Brain}, Phys,
Rev. Lett.  \textbf{91}, 078101 (2003).

\bibitem{yu05} Yu Y., R. Romero and T.S. Lee,  \emph{Preference of Sensory Neural Coding for 1/f Signals}, \textit{Phys, Rev. Lett}. 
\textbf{94}, 108103 (2005).

\bibitem{gong07} P. Gong, A.R. Nikolaev and C. van Leeuwen, \emph{Intermittent dynamics underlying the intrinsic fluctuations of the collective synchronization
patterns in electrocortical activity}, \textit{Phys.
Rev. E} \textbf{76}, 011904 (2007).

\bibitem{correll08} J. Correll, \textit{J. Person. Social Psychol}. \emph{1/f noise and effort on implicit measures of bias.}, \textbf{94}, 48 (2008).


\bibitem{gemignani} P.  Allegrini, D.  Menicucci, R.  Bedini,  L.  Fronzoni, A.  Gemignani, P.  Grigolini, B. J. West, P. Paradisi, \emph{Spontaneous brain activity as a source of ideal 1/f noise}, Phys. Rev. E, {\bf 80}, 061914  (2009).  

\bibitem{buiatti}  M. Buiatti, D. Papo, P.-M. Baudonni\`{e}re, Van Vreeswijk, \emph{Feedback modulates the temporal scale-free dynamics of brain electrical activity in a hypothesis testing task}, Neuroscience {\bf146}, 1400 (2007). 

\bibitem{heartbeat} P. Allegrini, P. Grigolini, P. Hamilton, Palatella, G. Raffaelli, \emph{Memory beyond memory in heart beating, a sign of a healthy physiological condition},
Phys. Rev. E, {\bf 65}, 041926 (2002). 

\bibitem{gyanendra}  G. Bohara, D. Lambert, B. J. West, P. Grigolini \emph{Crucial events, randomness and multi-fractality in heartbeats}, submitted to Phys. Rev. E. 

\bibitem{meditation} D. Kim, K. Lee, J. Kim,  M. Whang, S. W. Kang, \emph{Dynamic correlations between heart and brain rhythm during Autogenic meditation}, Front. Hum. Neurosci. {\bf 7}, 414  (2013). 









 \bibitem{pentland}  A. Pentland, \emph{To Signal Is Human},  American Scientist, { \bf 98}, 204 (2010). 





\bibitem{kello} D. H. Abney, Al.Paxton, R. Dale, C. T. Kello, \emph{Complexity Matching in Dyadic Conversation}, Journal of Experimental Psychology: General, {\bf 143}, 2304 (2014).  

\bibitem{iberall}  A. S. Iberall, \emph{A Physical (Homeokinetic) Foundation for the Gibsonian Theory of Perception and Action}, Ecological Psychology, {\bf 7}, 37 (1995). 


\bibitem{singlemolecules} M. C. Leake, \emph{The physics of life: one molecule at a time}, Phil.
Trans. R. Soc. B {\bf 368} ,  20120248 (2012). 

\bibitem{chemphyschem} S. Burov, J.-H. Jeon, R. Metzler, E. Barkai, \emph{Single particle tracking in systems showing anomalous diffusion:
the role of weak ergodicity breaking}, Phys. Chem. Chem. Phys., {\bf13}, 1800 (2011). 

\bibitem{stephen} D. G. Stephen, J. A. Dixon, \emph{Strong anticipation: Multifractal cascade dynamics modulate scaling in synchronization behaviors},Chaos, Solitons \& Fractals, {\bf 44}, 160 (2011).  

  

\bibitem{challenge}  P. Jizba, J.  Korbel,   \emph{Multifractal diffusion entropy analysis: Optimal bin width of probability histograms}, Physica A, {\bf  413 } , 438 (2014) .

\bibitem{feder88} J. Feder, \textit{Fractals}, Plenum Press, New York (1988).


\bibitem{ikeda1}  K. Ikeda, K. Matsumoto.\emph{High-dimensional chaotic behavior in
systems with time-delayed feedback},  Physica D, {\bf 29}, 223 (1987). 

\bibitem{ikeda2}
K. Ikeda, H. Daido, O. Akimoto, \emph{Optical turbulence: chaotic behavior of transmitted light from a ring cavity},  Phys. Rev. Lett. {\bf 45}, 709 (1980). 

\bibitem{voss} H. U. Voss, \emph{Anticipating chaotic 
syncronization}, Phys. Rev. E, {\bf 61}, 5115 (2000). 

\bibitem{temporalcomplexity} M.  Turalska, B. J. West  and Paolo Grigolini, \emph{Temporal complexity of the order parameter at the phase transition}, Phys. Rev. E {\bf 83}, 061142 (2011).

\bibitem{zare} M. Zare, P. Grigolini, \emph{Cooperation in neural systems: Bridging complexity and periodicity}, 
Phys.  Rev E {\bf 86}, 051918 (2012). 


\bibitem{original} S. Bianco, E. Geneston,  P. Grigolini and M. Ignaccolo,  \emph{Renewal  aging as emerging property of phase synchronization},  Physica A {\bf 387}, 1387 (2008). 

\bibitem{second} M. Turalska, M.  Lukovic, B. J. West, Paolo Grigolini, \emph{Complexity and Synchronization}, Phys. Rev. E {\bf 80}, 021110, 1-6, (2009).

\bibitem{third} M. Turalska, B. J. West, P. Grigolini, \emph{Temporal complexity of the order parameter at the phase transition}, Phys. Rev. E {\bf 83}, 061142 (2011).

\bibitem{fourth} N. W. Hollingshad, M. Turalska, P. Allegrini, Br. J. West, P. Grigolini,  \emph{A new measure of network efficiency}, Physica A , {\bf 391}, 1894 (2012). 

\bibitem{fifth}  M. Turalska, E. Geneston, B. J. West, P. Allegrini, P. Grigolini, \emph{Cooperation-induced topological complexity: a promising road to fault tolerance and Hebbian learning}, Frontiers in Fractal Physiology, {\bf 3}, 52, 1-7 (2012). 
\bibitem{20} B. J. West, M. Turalska, P. Grigolini, \emph{Networks of Echoes: Imitation, Innovation and Invisible Leaders}, Springer International (2014). 

\bibitem{detrended}  J. W. Kantelhardt, S. A. Zschiegner,
E. Koscielny-Bunde, S. Havlin,  A. Bunde,
H. E. Stanley, \emph{Multifractal detrended  fluctuation analysis of nonstationary time series}, Physica A {\bf 316}, 87 (2002).

\bibitem{peng} C.-K. Peng, S.V. Buldyrev, S. Havlin, M. Simons, H.E. Stanley, A.L. Goldberger, Phys. Rev. E {\bf 49}, 1685 (1994).

\bibitem{timedelay} M. Lukovi\'{c}, F. Vanni, A. Svenkeson,  \emph{Transmission of information at criticality}, Physica A {\bf 416}, 430 (2014).  

\bibitem{sotc} K. Mahmoodi, B. J. West, P. Grigolini, \emph{Self-Organizing Complex Networks: individual versus global rules}, submitted to Frontiers.

\bibitem{levyflight} N. Piccinini, B. J. West, P. Grigolini, \emph{Transfer of information from one to another complex network: How to bypass the technical and theoretical problems raised by criticality-induced ergodicity breaking?}, submitted to Phys. Rev. 
































\bibitem{aging} P. Allegrini, F. Barbi, P. Grigolini, and P. Paradisi, \emph{Renewal, modulation, and superstatistics in times series}
Phys. Rev. E
{\bf 73}, 046136 (2006).





\bibitem{cognition} F. Vanni, M. Lukovi\'{c}, P. Grigolini,
\emph{Criticality and Transmission of Information in a Swarm of Cooperative Units}, Phys. Rev. Lett, {\bf 107}, 07813 (20111). 


\bibitem{failla} R. Failla, P. Grigolini,M. Ignaccolo, A. Schwettmann, \emph{Random growth of interfaces as a subordinated process}, Phys. Rev.  E {\bf 70}, 010101(R) (2004).

\bibitem{rohisha} 
K.  Mahmoodi ,B. J.West, P. Grigolini1\emph{Self-organizing Complex Networks: individual versus global rules}, Frontiers in Physiology, {\bf 8}, 478 (2017). 





 
 
 
 
 \end{references}
\end{document}